\documentclass[letter]{aa} 
\usepackage{graphicx}
\usepackage{txfonts}
\usepackage{natbib}
\bibpunct{(}{)}{;}{a}{}{,}
\begin{document}
   \title{Dust-grain processing in circumbinary discs around evolved binaries.
         The RV\,Tauri spectral twins RU\,Cen and AC\,Her.
\thanks{Based on observations obtained at the European Southern Observatory (ESO),
La Silla, observing program 072.D-0263, on observations made with the 1.2\,m Flemish Mercator telescope
at Roque de los Muchachos, Spain, the 1.2\,m Swiss Euler telescope at La Silla,
Chile and on observations made with the Spitzer Space Telescope (program id 3274), which is operated 
by the Jet Propulsion Laboratory, California Institute of Technology under a contract with NASA.}
}

   \author{
          C. Gielen\inst{1}
          \and
          H. Van Winckel\inst{1}
          \and
          L.B.F.M. Waters\inst{1,2}
          \and
          M. Min\inst{2}
          \and
          C. Dominik\inst{2}
          }


   \institute{Instituut voor Sterrenkunde,
              Katholieke Universiteit Leuven, Celestijnenlaan 200D, 3001 Leuven, Belgium\\
	      \email{Clio.Gielen@ster.kuleuven.be}
              \and
              Sterrenkundig Instituut 'Anton Pannekoek', 
              Universiteit Amsterdam, Kruislaan 403, 1098 Amsterdam, The Netherlands\\
            }

   \date{Received ; accepted }
  \abstract
   {We study the structure and evolution of the circumstellar discs around evolved binaries and their impact on the evolution of the central
   system.}
   {By combining a wide range of observational data and techniques, we aim to study in detail the binary nature of RU\,Cen and AC\,Her, as well as the
  structure and mineralogy of the circumstellar environment.}
   {We combine a multi-wavelength observational program with a detailed 2D radiative transfer study. Our radial velocity program is
  instrumental in the study of the nature of the central stars, while our Spitzer spectra complimented with the broad-band spectral
  energy distribution (SED) are used to constrain mineralogy, grain sizes and physical structure of the circumstellar environment.}
   {We determine the orbital elements of RU\,Cen showing that the orbit is highly eccentric with a large velocity amplitude despite 
  the rather long period of 1500 days. The infrared spectra of both objects are very similar and the spectral dust
  features are dominated by magnesium-rich crystalline silicates. The small peak-to-continuum ratios are
  interpreted as being due to large grains. Our model contains two components with a cold midplain dominated by large grains, and
  the near- and mid-IR which is dominated by the emission of smaller silicates. The infrared excess is well modelled assuming a hydrostatic passive irradiated
  disc. The profile-fitting of the dust resonances shows that the grains must be very irregular.}
   {These two prototypical RV\,Tauri pulsators with circumstellar dust are binaries where the dust is trapped in a stable disc. The
  mineralogy and grain sizes show that the dust is highly processed, both in crystallinity and grain size. The cool crystals show that
  either radial mixing is very efficient and/or that the thermal history at grain formation has been very different from that in
  outflows. The physical processes governing the structure of these discs are very similar to those observed in protoplanetary discs
  around young stellar objects.}
   \keywords{stars: AGB, post-AGB -
             stars: evolution -
             stars: binaries -
             stars: circumstellar matter -
             stars: individual: RU\,Cen -
             stars: individual: AC\,Her
               }
  \titlerunning{The RV\,Tauri spectral twins RU\,Cen and AC\,Her.}
   \maketitle
%

\section{Introduction}
\label{introduction}

The RV\,Tauri class of objects contains highly luminous stars showing
large-amplitude photometric variations with alternating deep and shallow
minima. The members are located in the high luminosity end of the population\,II
instability strip and the photometric variations are interpreted
as being due to radial pulsations.
There are two different photometric classes: the RVa stars are objects
with a constant
mean magnitude while the RVb objects display a long-term variation in
their mean magnitude.
\citet{preston63} introduced a spectroscopic classification of the
RV\,Tauri stars, using unfortunately the same alphabetic letters for
the naming: RVA objects show strong absorption lines,
RVB objects are of a somewhat hotter spectral type but are weak lined with
enhanced CN and CH bands. The RVC objects are also weak lined but show no enhanced CN and CH molecular
band strength.
A significant fraction of the RV\,Tauri stars show a large IR excess
due to circumstellar dust and \citet{jura86} identified them as post-AGB objects
on the basis of this IR excess, their luminosities and mass-loss history.

The photospheric content of RV\,Tauri stars is, however, very
different from what could be expected in post-3rd dredge-up objects: they do
not show high C-abundances or s-process overabundances
but instead often show a depletion pattern in their photospheres
\citep{giridhar94,giridhar98,giridhar00,gonzalez97b,gonzalez97a, vanwinckel98, maas05}.
This abundance pattern is the result of gas-dust separation followed by reaccretion
of the gas, which is poor in refractory elements.
\citet{waters92} proposed that the most likely circumstance for this
process to occur is when the dust is trapped in a circumstellar disc.
This depletion phenomenon is also observed in binary post-AGB stars with a
disc \citep{vanwinckel95}. This led \citet{vanwinckel99} to
suggest that the depleted RV\,Tauri stars must also be binaries with a disc.
The likely presence of a Keplerian circumstellar disc was further
shown by the systematic study by \citet{deruyter05,deruyter06} of
a large sample of binary post-AGB and RV\,Tauri stars.

\citet{blocker95} list the post-AGB evolutionary tracks of
single stars of different initial mass. Evolutionary tracks
for binary post-AGB stars have not yet been determined, however, since we find evidence
that these stars have been shortcut on their AGB evolution (Sect. \ref{conclusion}), 
we expect longer lifetime scales because of the expected lower core
masses. Typical post-AGB lifetimes of both stars are estimated to be of the order of $\sim 10^4\,$yr.

To investigate the special evolutionary status of RV\,Tauri stars and
to research in detail the interplay between the photospheric and
circumstellar environment in these systems, we focus in this paper on
two well known RV\,Tauri stars: AC\,Her and RU\,Cen.

AC\,Her has been extensively studied in the literature. It is a binary
RV\,Tauri star of photometric class a, with an orbital period of 1200
days \citep{vanwinckel98}. It is a very regular pulsator with a
formal pulsation period (timespan between two successive
deep photometric minima) of 75.5 days \citep{zsoldos93}. The mean
magnitude of AC\,Her is $m_V=7.69$ mag and the amplitude $\bigtriangleup m_V=2.31$
mag \citep{lloydevans85}. The presence of two strong shock waves in
every formal pulsation cycle, causing line-profile deformations in the
spectra of AC\,Her and R\,Scuti has been discussed by
\citet{gillet89,gillet90}. AC\,Her shows a chemical depletion pattern
\citep{vanwinckel98,giridhar00} that is attributed to the presence of
a stable Keplerian dusty disc. The presence of such a disc has also
been proposed by \citet{jura99} who interpret the detection of weak CO
rotational emission lines with a small velocity width
\citep{bujarrabal88,jura95} as a signature of such a long-lived dust
reservoir. The presence of highly crystalline silicates in the
infrared spectrum \citep{molster99} and the strong millimeter
continuum flux from large dust grains \citep{shenton95,jura99} further
corroborate this conclusion. \citet{jura00} claimed to have resolved
the circumstellar material around AC\,Her using N and Q-band
imaging. \citet{close03} however detect no significant extended
structure around AC\,Her, using adaptive optics at mid-infrared
wavelengths with a higher spatial resolution.

RU\,Cen also is an RV\,Tauri star of spectroscopic class B and photometric
class a. It is a regular pulsator with a pulsation period of 64.6
days, a mean magnitude of $m_V=9.05$ mag and an amplitude of
$\bigtriangleup m_V=1.28$ mag \citep{pollard96}. During every formal pulsation cycle
(the period between two photometric deep minima) two shock waves
propagate through the atmosphere \citep{maas02}. The same paper reports the detection of a significant
radial velocity variation on a longer time scale and attributes this to
be due to orbital motion. No orbit could be determined, however.

\begin{table}
\caption{The name, equatorial coordinates $\alpha$ and $\delta$ (J2000), the effective temperature T$_{eff}$,
the surface gravity $\log g$ and the metallicity [Fe/H] of the programme stars RU\,Cen and AC\,Her.
Model parameters are taken from \citet{maas02} and \citet{vanwinckel98}. Typical errors are $\Delta$\,T$=250$\,K,
$\Delta \log g=0.5$. }
\label{modelparameters}
\centering
\begin{tabular}{cccccc}
\hline \hline
 name & $\alpha$ (J2000) & $\delta$ (J2000) & T$_{eff}$ & $\log g$ & [Fe/H] \\ & (h m s) & ($^\circ$ ' '') & (K) & (cgs) & \\
\hline RU\,Cen & 12 09 23.7 & -45 25 35 & 6000 & 1.5 & -2.0 \\ AC\,Her
& 18 30 16.2 & +21 52 00 & 5500 & 0.5 & -1.5 \\ \hline
\end{tabular}
\end{table}

Both stars are very regular RV\,Tauri pulsators with very similar chemical
depletion patterns, atmospheric parameters and pulsational
stability.
In this paper we report our detailed comparative study of both objects based on
our optical monitoring and Spitzer Space Telescope spectra.  The
outline of the paper is as follows: in Sect. \ref{observations} we
give an overview of the different observations and reduction
strategies. Sect. \ref{sed} contains the construction of the spectral
energy distributions and colour excess determination.
In Sect. \ref{binarity} we discuss the spectral monitoring
and deduce the binary model of RU\,Cen. The analysis of the infrared
spectra and the spectral fitting is done in Sect. \ref{analysis}. In
Sect. \ref{sedfit} we model the observed SEDs using a passive disc
model. The discussion of our different results and our conclusions are
presented in Sect. \ref{discussion} and Sect. \ref{conclusion}.

\section{Observations and reduction}
\label{observations}

\subsection{Spitzer}
We used the Infrared Spectrograph (IRS) aboard the Spitzer Space Telescope
in February 2005 to obtain high- and low-resolution spectra of
RU\,Cen. The spectra were observed with combinations of the
short-low (SL), short-high (SH), long-low (LL) and long-high (LH)
modules. SL ($\lambda$=5.3-14.5\,$\mu$m) and LL
($\lambda$=14.2-40.0\,$\mu$m) spectra have a resolving power of
R=$\lambda/\bigtriangleup\lambda \sim$ 100
and the SH ($\lambda$=10.0-19.5\,$\mu$m) and LH
($\lambda$=19.3-37.0\,$\mu$m) spectra have a resolving power of $\sim$ 600.

The spectra were extracted from the SSC raw data pipeline version S13.2.0 products,
using the c2d Interactive Analysis reduction software package
\citep{kessler06,lahuis06}. This data
processing includes bad-pixel correction, extraction, defringing and order
matching. Individual orders are corrected for offsets, if necessary, by small scaling
corrections to match the bluer order.

\subsection{TIMMI2}
We completed our infrared spectra of RU\,Cen with spectra in the 10\,$\mu$m
region, taken with the TIMMI2 instrument
mounted on the 3.6\,m telescope at the ESO La Silla Observatory in March
2004. The low-resolution ($ R \sim$ 160) N band grism
was used in combination with a 1.2\,arcsec slit, the pixel scale in the
spectroscopic mode of TIMMI2 is 0.45\,arcsec.
For the reduction of the spectra we used the method described in \citet{vanboekel05}.

\subsection{CORALIE}

We extended the radial velocity monitoring reported by \citet{maas02}
with new data obtained with the same spectrograph CORALIE attached to
the same 1.2\,m Swiss Euler telescope. In total we accumulated 151 raw spectra
between June 2000 and July 2006 with a typical sampling of 3 runs
of 10 days spread over every semester. The radial velocity was determined
by off-line cross-correlation using a spectral mask tuned to the
spectral properties of RU\,Cen \citep{maas02}. The internal error for
every measurement was quantified by the standard deviation of a
50-point bisector through the cross-correlation profile. The bisector
was determined on an equidistant sampling, starting from 2 times the width ($\sigma$) of the
Gaussian fit through the cross-correlation profile down to the
minimum. From January 2005 onwards, we used the HARPS-software release
to determine the cross-correlation profile. We extended the baseline
by including the radial velocity data of \citet{pollard97} which are
based on high-quality high-resolution spectra.

\section{SED determination}
\label{sed}

We collected photometric data so as to construct the spectral energy
distributions (SEDs) homogeneously \citep{deruyter05,deruyter06}. The main problem in constructing
such SEDs of pulsating stars with large amplitudes and often strong
cycle-to-cycle variability is that equally phased data are necessary.
Since we do not have these data available, photometric maxima were
used for the SED construction.

Our SED construction gives E(B-V)=0.4$\pm$0.3 for RU\,Cen and
E(B-V)=0.2$^{+0.3}_{-0.2}$ for AC\,Her.
The colour excess E(B-V) determination method was adopted from \citet{deruyter06}.
We estimate total extinction by dereddening observed photometry, using the average extinction
law of \citet{savage79}. Minimising the difference between the dereddened observed fluxes and the
adopted Kurucz model gives total colour excess E(B-V).  
Model parameters for our programme stars (Table\ \ref{modelparameters}) are taken from \citet{maas02} and \citet{vanwinckel98}.

The SEDs of RU\,Cen and AC\,Her have also
been discussed in \citet{maas02} and \citet{deruyter06} who find a
total reddening for RU\,Cen of respectively E(B-V)=0.6$\pm$0.1 and
E(B-V)=0.3$\pm$0.3 and a total reddening for AC\,Her of E(B-V)=0.1$^{+0.3}_{-0.2}$.

For both stars we find a broad infrared excess, starting around L-M. We find
a value for
the energy ratio $L_{IR}/L_\ast\approx0.15$ for
RU\,Cen and $L_{IR}/L_\ast\approx0.35$ for AC\,Her.

\section{Binary orbit of RU\,Cen}
\label{binarity}

Together with AC\,Her, RU\,Cen is known to be one of the most regular
RV\,Tauri star pulsators. It has a stable formal period of 64.60 days  and a total
peak-to-peak amplitude of 1.3 magnitudes in V
\citep{pollard97}. There is no long-term photometric modulation
detected.

\begin{figure}
\vspace{0cm}
\hspace{0cm}
\resizebox{8cm}{!}{\rotatebox{90}{\includegraphics{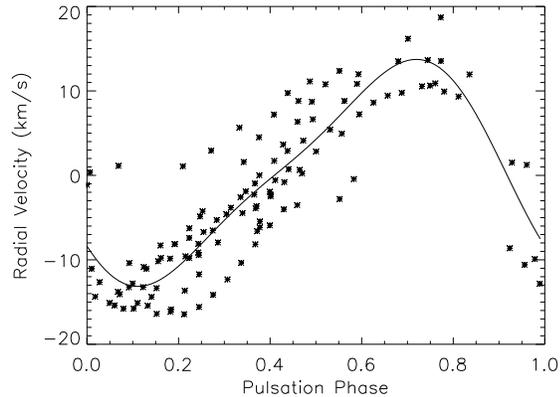}}}
\caption{The pulsational modulation of the radial velocity of RU\,Cen
  after cleaning of the orbit. The full line is the harmonic fit (one
  overtone included) with the pulsation period of 32.36
  days. Zero phase was taken arbitrarily on the first datapoint
  obtained (JD2448388.93). The number of pulsation cycles covered is
  170. The fractional variance reduction of the fit is 77\%. Note the
  very high pulsational amplitude.}
\label{fig:pulsmodel}%
\end{figure}

\begin{figure}
\vspace{0cm}
\hspace{0cm}
\resizebox{8cm}{!}{\includegraphics{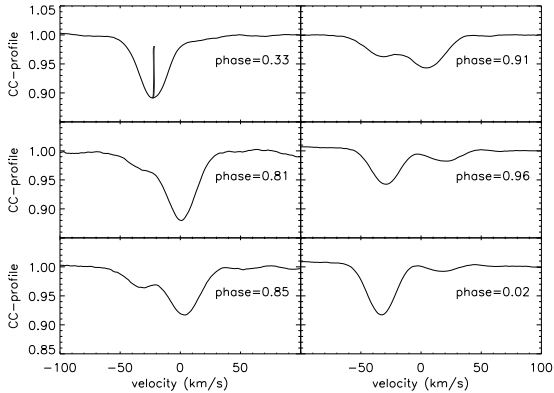}}
\caption{The cross-correlation profiles obtained at different
pulsational phases. The upper left panel gives the cross correlation
profile at phase 0.33 with the 50-point bisector which is used to
quantify the radial velocity. In the other panels, we witness the
propagation of a strong shock in the lineforming region of the
atmosphere.}
\label{fig:pulscc}%
\end{figure}

The pulsational modulation of the radial velocity
is very significant (see Fig.~\ref{fig:pulsmodel}) making orbital
detection far from straightforward. Any cycle-to-cycle variability
will make that a systematic cleaning of the pulsation from the raw radial velocity data
yields a residual. Moreover, strong atmospheric shocks
associated with the RV\,Tauri pulsations passing through the line-forming
region \citep[e.g.][]{gillet90}, have a strong effect on the
line-profiles.
 This makes the determination of the stellar radial velocity at those
pulsational phases problematic. In Fig.~\ref{fig:pulscc}\, we show a few
cross-correlation profiles at different phases in the pulsation
cycle. The propagation of the shock is well illustrated and in the case of
RU\,Cen the non-linear behaviour leads to a very significant drop in
velocity of more than 20 km\,s$^{-1}$ over a small phase interval.
The shocks in RU\,Cen are so energetic that
during these phases, He lines are observed in emission \citep{maas02}.

Despite the strong pulsational modulation in the radial velocity data,
variability in the radial velocity is detected with a much longer
time scale. We interpret this as being due to orbital motion and modelled this with a Keplerian
model of a binary star.

To obtain the orbital elements, we only retained the high quality data
outside the pulsation phases where the strong shock is visible
in the cross-correlation profile. To do so, we required that in all
data used for the orbital detection, the 50-point bisector has a variance of
less than 1 km\,s$^{-1}$.
We performed an iterative process on the raw
velocity data in which we
cleaned the orbital solution from the raw radial velocity data to
obtain a good model of the pulsation cycle itself. We used PDM (Phase
Dispersion Minimalisation method developed by \citealt{stellingwerf78})
to quantify the fundamental pulsation period (time between successive
deep and shallow photometric minimum) and determined a harmonic fit
with one overtone as a model description of the pulsation.
We then cleaned the raw radial velocity data by the
pulsation model and performed the next least-square fit of the orbit.
We stopped the iteration when the changes in the orbital parameters
became less than the error.

The final result is that we indeed found an  orbital solution with a
period of 1489 $\pm$ 4 days. The errors given in the table are the formal errors
obtained using the covariance matrix \citep{hadrava04}.
The mass function is 0.83 M$_{\odot}$ and the semi-major axis is
a$_{1} \, \sin i$ = 2.4 AU.

\begin{table}
\caption{The orbital elements of the binary RU\,Cen. P is the orbital period, JD the periastron passage, K gives the semi-amplitude,
  e the eccentricity, $\omega$ is the longitude of periastron, $\gamma$ is the system radial
  velocity, a$_{1}$ gives the semi-major axis of the primary and $i$ is the inclination.}
\begin{center}
\begin{tabular}{cccc}
\hline
\hline
Element                &    Value   &  Formal Error (1$\sigma$) & Unit  \\
\hline
P                      &   1489     &    4           &  days \\
JD$_{\rm periastron}$  &   2451378  &    10          &  days  \\
K                      &   21.9     &    0.7         &  km\,s$^{-1}$ \\
e                      &   0.60     &    0.03        &        \\
$\omega$               &   133      &    4           &  degrees\\
$\gamma$               &   -28.4    &    0.4         &  km\,s$^{-1}$ \\
rms                    &    4.5     &                &   km\,s$^{-1}$
\\
Mass Function         &   0.83     &                 & M$_{\odot}$ \\
a$_{1}$\, $\sin i$    &   2.4      &                 & A.U. \\

\hline
\end{tabular}
\end{center}
\end{table}

\begin{figure}
\vspace{0cm}
\hspace{0cm}
\resizebox{8cm}{!}{\rotatebox{90}{\includegraphics{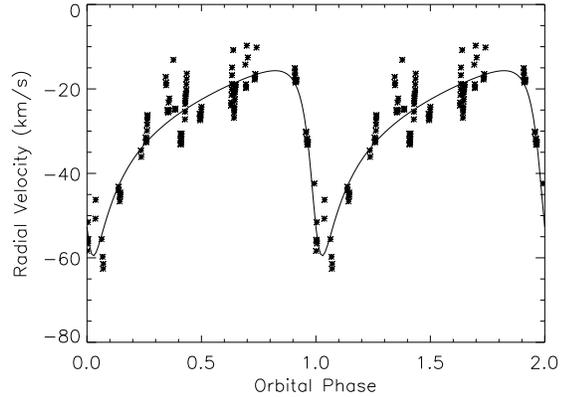}}}
\caption{The radial velocity data of RU\,Cen, cleaned for the
  pulsational modulation and folded on the orbital period of 1489
  days. The epoch of zero phase is the periastron passage
  (JD2451378). The full line represents the  least square orbital
  solution discussed in the text. Note that for clarity the figure
  samples all data twice.}
\label{fig:orbit}%
\end{figure}

\section{Analysis infrared spectra}
\label{analysis}

\subsection{General}

\begin{figure}
\vspace{0cm}
\hspace{0cm}
\resizebox{8cm}{!}{ \includegraphics{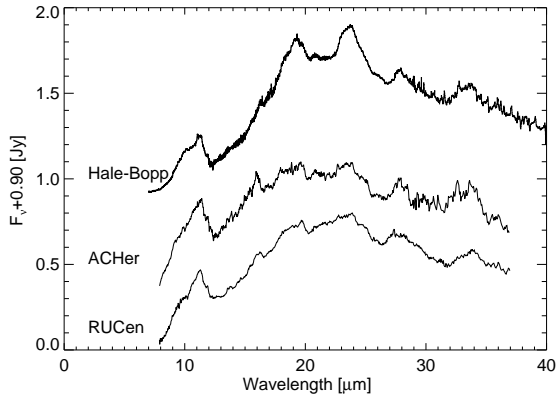}}
\caption{ISO-SWS spectra from the solar system comet Hale-Bopp \citep{crovisier97},
the RV\,Tauri star AC\,Her \citep{molster99} and
combined TIMMI2 and Spitzer-IRS spectrum of RU\,Cen. The spectra are
normalised and offset for comparison.
The dust emission features on top of the continuum are identified as being due
to enstatite and forsterite grains.}
\label{ondereen}%
\end{figure}

A first look at the infrared spectra of AC\,Her and RU\,Cen shows their
striking similarity (see Fig.\ \ref{ondereen}), both in the global
shape and in the dust emission features. In both spectra there is
a lack of a strong 10\,$\mu$m amorphous silicate feature, while the 20\,$\mu$m
amorphous feature is prominent. AC\,Her and RU\,Cen show strong emission
features around
11.3 - 16.2 - 19.7 - 23.7 - 28 - 33.6\,$\mu$m which we can identify as
features of forsterite and enstatite, two abundant silicate crystals. In
none of the spectra is there evidence for a carbon-rich component. Not only
are the infrared spectra of AC\,Her and RU\,Cen very similar to each other,
they also show a strong resemblance
to the infrared spectrum of the solar system comet Hale-Bopp \citep{bouwman03,min05b}.

\citet{min05b} model the thermal emission and degree of linear polarisation of radiation scattered
by grains in the coma of the comet Hale-Bopp.
The largest contribution in dust, about 75\% of total dust mass,
is made up of amorphous silicate grains, with dust sizes
from 0.01\,$\mu$m up to 93\,$\mu$m, and large amorphous carbon grains ($\sim$ 10\,$\mu$m). 
Small crystalline silicates make up only 5\% of the total
dust mass of Hale-Bopp but this is sufficient to have this strong
spectral signature in the IR spectrum.

\subsection{Feature identification}
\label{feat_ident}

The amorphous and crystalline features seen in the spectra of AC\,Her and
RU\,Cen are identified as caused by the most
commonly found dust species in the circumstellar environment \citep{molster02a,molster02b,molster02c,min07},
namely glassy and crystalline silicates with an olivine and pyroxene stoichiometry.
We further take the commonly used term ``amorphous and crystalline
olivine and pyroxene'' to describe these dust species. Amorphous olivine (Mg$_{2x}$Fe$_{2(1-x)}$SiO$_4$,
where 0$\leq$x$\leq$1 denotes the magnesium content) has very prominent
broad
features around 9.8\,$\mu$m and 18\,$\mu$m. These features (also called the
10\,$\mu$m and 20\,$\mu$m features) arise respectively from the Si-O
stretching mode and the O-Si-O bending mode. For large grains the
9.8\,$\mu$m feature gets broader and shifts to redder wavelengths. Amorphous
pyroxene
(Mg$_{x}$Fe$_{1-x}$SiO$_3$) shows a 10\,$\mu$m feature similar to that of
amorphous olivine, but shifted towards shorter wavelengths. Also the shape
of the
20\,$\mu$m feature is slightly different.
Crystalline olivine and pyroxene have very distinct emission features and
comparing with the features seen in the spectra of AC\,Her and RU\,Cen,
we conclude that the Mg-rich end members, forsterite (Mg$_2$SiO$_4$) and
enstatite (MgSiO$_3$), dominate our spectra.
Forsterite condenses directly from the gas-phase at high temperatures
($\approx 1500$\,K) or it may form by thermal annealing of amorphous
silicates, diffusing the iron out of the lattice-structure. Enstatite can
form in the gas-phase from a reaction between forsterite and silica, or it
may also form by a similar thermal annealing process as forsterite \citep{bradley83,tielens97}.

The observed spectra of AC\,Her and RU\,Cen show a shift from the amorphous
18\,$\mu$m feature towards 20\,$\mu$m when comparing with synthetic spectra
of amorphous olivine and pyroxene. This could point to the dominance of
Mg-rich amorphous dust, which also shows this shift to redder wavelengths.
Photospheric depletion in iron, which we detect in RU\,Cen and AC\,Her \citep{maas02,vanwinckel98},
can be understood when the iron is locked up in the circumstellar dust \citep{waters92}.
The lack of iron in the detected silicates is therefore
surprising. If both the crystalline and amorphous silicates are devoid of
iron, this could mean that iron is stored in metallic iron or iron-oxide
\citep{sofia06}. Metallic iron has no distinct features but still a
significant contribution in opacity, especially at shorter wavelengths,
making it very hard to detect directly.

\subsection{Profile fitting}
\label{profilefit}

\begin{figure}
\vspace{0cm}
\hspace{0cm}
\resizebox{9cm}{!}{ \includegraphics{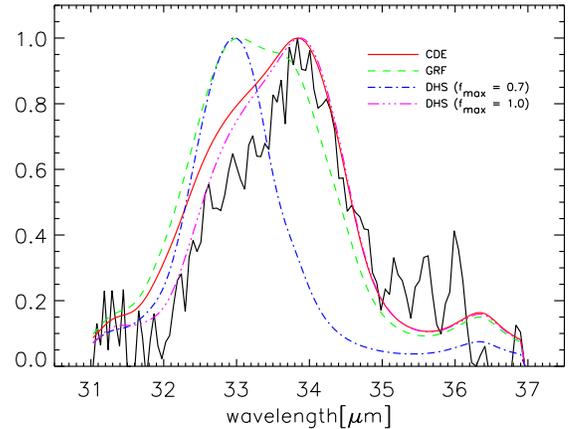}}
\caption{Normalised and continuum subtracted 33.6\,$\mu$m emission feature
of RU\,Cen together with mass absorption coefficients for different
forsterite shape distributions. The DHS forsterite grains both have grain sizes of
0.1\,$\mu$m. From the figure it is clear that the grains in DHS approximation must have
$f_{max}=1.0$. For this emission feature the difference between DHS grains of 0.1\,$\mu$m
and 1.5\,$\mu$m is minimal.}
\label{profiles1}%
\end{figure}

\begin{figure}
\vspace{0cm}
\hspace{0cm}
\resizebox{9cm}{!}{ \includegraphics{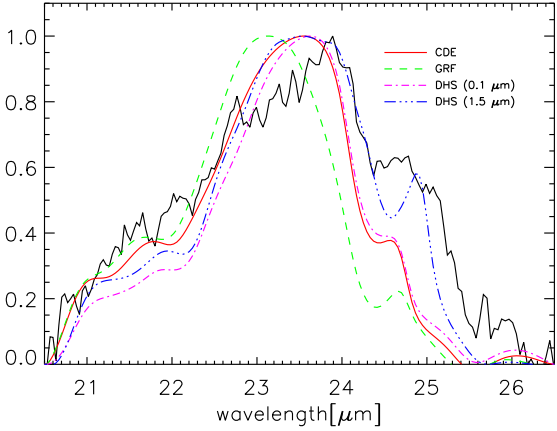}}
\caption{Normalised and continuum subtracted 23.7\,$\mu$m emission feature
of RU\,Cen together with mass absorption coefficients for different
forsterite shape distributions. The DHS forsterite grains both have $f_{max}=1.0$.
Grains of 1.5\,$\mu$m are necessary to fit the observed bump at 25\,$\mu$m.}
\label{profiles2}%
\end{figure}

Our aim is to fit the observed crystalline emission features of AC\,Her and
RU\,Cen with synthetic spectra of forsterite and enstatite.
The conversion from laboratory measured optical constants of dust to mass
absorption coefficients is not straightforward and is largely
dependent on the adopted size, shape, structure and chemical composition of
the dust \citep{min03,min05a}.
These different dust approximations result in very different emission
feature profiles.
The spectrum produced by homogeneous spherical particles is very different
from that produced by more irregular particles.
This difference is much larger than the difference between synthetic spectra computed
using approximations of different irregular particles \citep{min03}.
We have access to a large sample of mass absorption coefficients of various dust shapes and
sizes. The sample consists of forsterite and
enstatite in Mie approximation \citep{aden51,toon81}, CDE (continuous distribution of ellipsoids, \citealp{bohren83}),
GRF (Gaussian random field particles, e.g. \citealp{grynko03,shkuratov05}) and DHS (distribution of hollow
spheres, \citealp{min03,min05a}) grain shapes. The DHS shaped particles are further characterised by
the
fraction of the total volume occupied by the central vacuum inclusion, $f$,
over the range $0<f<f_{max}$. The value of $f_{max}$ reflects the degree of
irregularity of the particles \citep{min03,min05a}.
Mie theory is used to model homogeneous spherical compact grains, while
CDE, GRF and DHS particles are more irregular.
Cross sections in CDE and GRF are computed under the assumption that the grains
are in the Rayleigh limit (that the grains are much smaller than the
wavelength of radiation, thus smaller than 0.1\,$\mu$m). The different grain
sizes for Mie and DHS dust particles range from 0.1\,$\mu$m till 10\,$\mu$m.
These different grain sizes produce emission features at very different
central wavelengths (see Figs.\ \ref{profiles1} and \ref{profiles2}) and larger grains mainly
contribute to the dust continuum.

To keep the number of free parameters in our $\chi^2$ minimisation
reasonable, we first determined and then fixed the best forsterite dust opacity
description. We carefully compared the observed dust emission profiles of
RU\,Cen and AC\,Her with emission profiles of different adopted forsterite
shapes. From Figs.\ \ref{profiles1} and \ref{profiles2} it is clear that different dust shapes
result in very different emission profiles.
We calculated that the best fit is obtained when using small ($<$
0.1\,$\mu$m) forsterite particles in CDE approximation and big (1.5\,$\mu$m)
forsterite particles in DHS approximation with $f_{max}=1.0$. The identification of the best
enstatite dust species is not as straightforward since the enstatite
emission features are often blended with more prominent forsterite features.
We have opted to keep the same dust size distribution for forsterite as for enstatite
since this is physically more plausible.

\subsection{Full feature fitting}

Identifying the underlying dust continuum distribution is by no means
straightforward. We opted to extract the dust continuum by connecting the
local minima of the crystalline features. This method was used both for the
synthetic spectra of forsterite and enstatite and the observed spectra of
RU\,Cen and AC\,Her to allow for a quantative comparison.
The best model fit was then determined using a $\chi^2$ minimalisation. Free
parameters are dust species, the fraction of the given dust species, dust
temperatures and the dust fractions at a given dust temperature.
This allows us to study the contribution of different dust temperatures and
the different dust species to the infrared spectral features.

\noindent The model emission profiles are then given by $$F_\lambda \sim
(\sum_i \alpha_i\kappa_i)*(\sum_j \beta_j B_\lambda(T_j))$$ where $\kappa_i$
is the mass absorption coefficient of dust component $i$ and $\alpha_i$
gives the fraction of that dust component, $B_\lambda(T_j)$ denotes the
Planck function at temperature $T_j$ and $\beta_j$ the fraction of dust in
that given temperature.
In this approach we assume that different grain species and grain sizes can
have equal temperatures and that the observed flux originates from an
optically thin region.

\begin{table}
\caption{Dust parameters deduced from our spectral fitting for RU\,Cen and AC\,Her.
Listed are the two dust temperatures, the fraction of dust with that given temperature
and the different fractions of dust in the given dust species. Small forsterite/enstatite
dust consists of CDE dust shapes with grain sizes $< 0.1\,\mu$m, large forsterite/enstatite dust consists
of DHS dust shapes of 1.5\,$\mu$m sized dust particles with $f_{max}=1.0$.}
\label{fitparameters}
\centering
\begin{tabular}{lcc}
\hline
\hline
 & RU\,Cen & AC\,Her \\
\hline
T$_1$=dust temperature 1 (K) & 150 & 100 \\
T$_2$=dust temperature 2 (K) & 600 & 800 \\
$\beta_1$=fraction dust in T$_1$ & 0.40 & 0.60 \\
$\beta_2$=fraction dust in T$_2$ & 0.60 & 0.40 \\
$\alpha_1$=fraction small forsterite  & 0.40 & 0.50 \\
$\alpha_2$=fraction large forsterite  & 0.30 & 0.20 \\
$\alpha_3$=fraction small enstatite   & 0.00 & 0.10 \\
$\alpha_4$=fraction large enstatite   & 0.30 & 0.20 \\
\hline
\end{tabular}
\end{table}

\begin{figure}
\vspace{0cm}
\hspace{0cm}
\resizebox{9cm}{!}{ \includegraphics{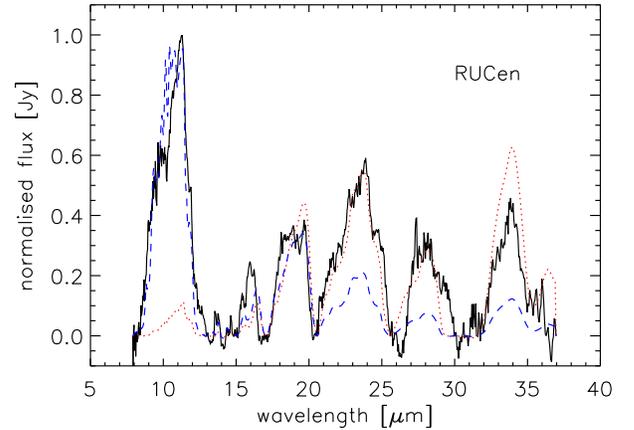}}
\caption{Continuum subtracted and normalised spectrum of RU\,Cen.
Overplotted two models of a forsterite-enstatite mixture at different
temperatures.
The dashed line represents a model at 150\,K and the dotted line a model at
600\,K. It is clear that a single-temperature model is insufficient to
reproduce
the spectrum of RU\,Cen.}
\label{2temp}%
\end{figure}

\begin{figure}
\vspace{0cm}
\hspace{0cm}
\resizebox{9cm}{!}{ \includegraphics{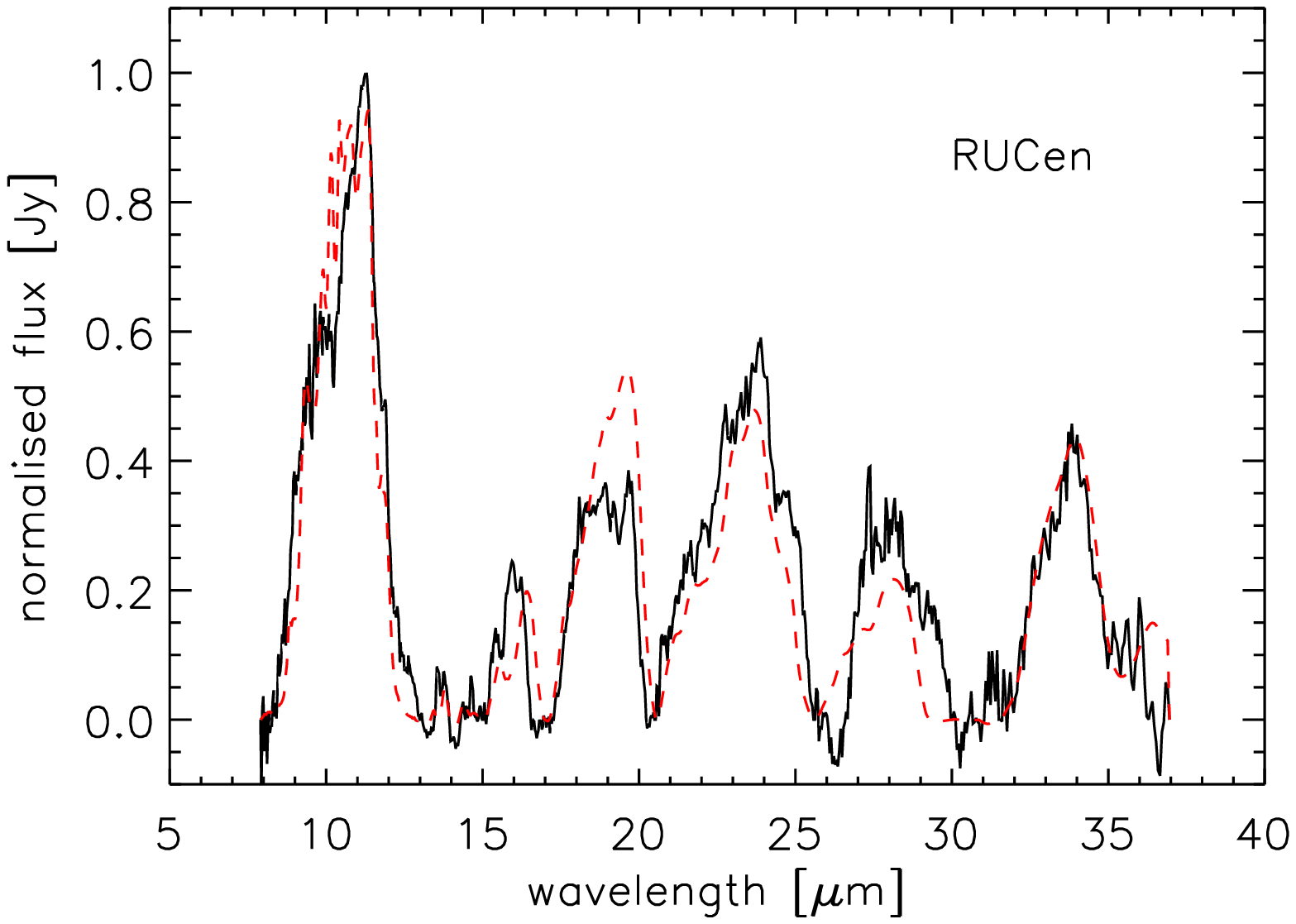}}
\caption{Best model fit to the continuum subtracted and normalised spectrum
of RU\,Cen. Fit parameters are given in Table\ \ref{fitparameters}.}
\label{modelRUCen}%
\end{figure}

\begin{figure}
\vspace{0cm}
\hspace{0cm}
\resizebox{9cm}{!}{ \includegraphics{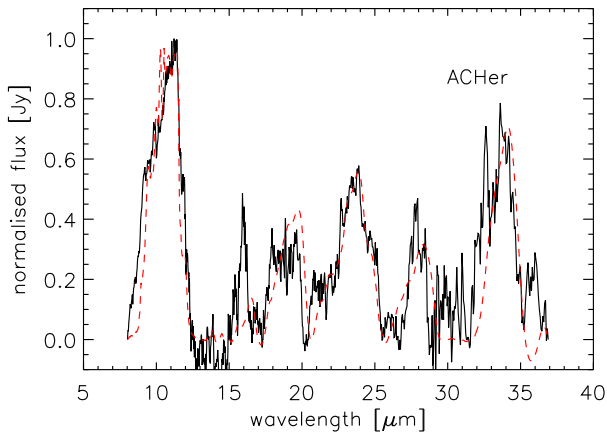}}
\caption{Best model fit to the continuum subtracted and normalised spectrum
of AC\,Her. Fit parameters are given in Table\ \ref{fitparameters}.}
\label{modelACHer}%
\end{figure}

From Sect. \ref{profilefit} we find that the best fit is obtained using very irregular grains, which was also
found in recent studies of dust in comets and protoplanetary discs
\citep[e.g.][]{crovisier97,bouwman01,bouwman03}.
As a next step, the dust species are kept fixed and the free parameters in the
$\chi^2$ minimalisation are dust temperatures, temperature fractions and
dust species fractions.

It is clearly not possible to fit the observed crystalline dust spectral features
of  RU\,Cen and AC\,Her with only one dust temperature (Fig.\ \ref{2temp}). 
The range of temperatures needed is, however, limited
and good fits can be achieved by allowing only two different dust
temperatures between 50\,K and 1000\,K  for
the adopted dust features (small/big forsterite/enstatite). An increase in
the number of dust temperatures yielded only a minor improvement in the
$\chi^2$ minimisation.

Results of our model fit are given in Table\ \ref{fitparameters} and Figs.\
\ref{modelRUCen} and \ref{modelACHer}. We find that both hot and cool
dust are necessary to reproduce the observed spectra and that AC\,Her and
RU\,Cen have a similar temperature distribution.
The hot dust temperature is less constrained and similar fits could be derived
with a temperature a few hundreds of Kelvin higher.
All our best models give a fraction of big enstatite grains that is higher
than the fraction of small enstatite grains, sometimes the fraction of small
grains is even zero. One explanation for the need of a high fraction of large grains
is the rather broad 11.3\,$\mu$m feature, which can only be fitted using a
considerable fraction of large grains. Including a contribution of amorphous
silicates in this broad feature makes the crystalline emission feature much
narrower and more peaked around 11.3\,$\mu$m, which is distinctive for small
forsterite grains. Of all the different observed
emission features, the 11.3\,$\mu$m feature is most sensitive to adopted grain size.
\citet{molster02b} already found for AC\,Her that the 11.3\,$\mu$m feature is well
fitted by only crystalline silicates but that an amorphous contribution
could not be excluded.
Including an amorphous component would reduce the 11.3\,$\mu$m feature and result 
in a larger fraction of small forsterite grains
and a reduction in big enstatite grains. The maximum dust temperature of the forsterite-enstatite mixture
decreases from temperatures around 700\,K to around 500\,K.

Our best models do not always match the observed features: in our modelled RU\,Cen
spectrum the 19\,$\mu$m feature is stronger than the 23\,$\mu$m feature,
while this is not the case in the observed spectrum of RU\,Cen. This
problem could arise from a problem in the data reduction, since at around 20\,$\mu$m,
getting a good order overlap of the different echelle orders proved to be problematic. Also the shape
of the 27\,$\mu$m feature is quite different in observed and modelled
spectra. Another puzzling fact is that the 16.2\,$\mu$m forsterite feature
seems to be shifted to the left and it is surprisingly strong in AC\,Her.
This is not a temperature effect and it may call for the inclusion of another mineral since
in RU\,Cen this feature is not well reproduced either.

\section{SED fitting disc model}
\label{sedfit}

\subsection{Method}

Broad-band SEDs are notoriously degenerate but together with the Spitzer
infrared spectral information, the circumstellar
physical characteristics are much better constrained. A firm first conclusion is that all spherical models
failed to fit both the SED and the infrared spectral data. Moreover, spherical
outflow models which fit the SED have evolutionary timescales which are much too
short compared to any evolutionary track in which a post-AGB star of spectral
type F is involved. We therefore concentrate on constructing detailed 2D disc
models.

\subsubsection{2D disc model}
\label{discmodel}

As a next step we performed an SED-fitting using a Monte Carlo code, assuming 
2D-radiative transfer in a passive disc model \citep{dullemond01,dullemond03,dullemond04}.
This code computes the temperature structure and density of the disc. The
vertical scale height of the disc is computed by an iteration process,
demanding vertical hydrostatic equilibrium. The dust grain property distribution
is fully homogeneous, and although this model can reproduce the SED,
dust settling timescales indicate that settling of large grains to the midplane occurs
and thus that an inhomogeneous disc model is necessary. 
Large grains are necessary to account for the
850\,$\mu$m flux in AC\,Her. As we do not possess a 850\,$\mu$m
fluxpoint for RU\,Cen, we estimated the 850\,$\mu$m flux by assuming a
blackbody slope from the IRAS 60\,$\mu$m flux redwards, as is observed in AC\,Her.
This submillimeter data is invaluable to constrain grain sizes in the disc.
In all other similar sources sampled \citep{deruyter06} the 850\,$\mu$m flux shows that
the flux is at the Rayleigh-Jeans slope connecting the 60\,$\mu$m IRAS flux point.

Dust settling time for a grain to migrate from height $z_0$ to $z$ is calculated using 
$$t_{set}=\frac{\pi}{2}\frac{\Sigma_0}{\rho_d a}\frac{1}{\Omega_k}\ln\frac{z}{z_0}$$ with $\Sigma_0$ the surface density, $\rho_d$ the particle
density, $a$ the grain size and $\Omega_k=\sqrt{\frac{GM_*}{r^3}}$ the Kepler rotation rate \citep{miyake95}. 
From our SED modelling (Sect. \ref{results}) we find that the surface density of these discs approximately is
$\Sigma=\Sigma_0 (\frac{R}{1\,AU})^{-0.5}$, with $\Sigma_0 \approx 0.2$\,gcm$^{-2}$.
This estimation means that grains larger than $100\,\mu$m can descend a distance of 50\,AU in less than 
$4\times10^5$ years. Grains of 850$\,\mu$m will even travel this distance in less than $6\times10^4$ years. 
This is similar to the estimated lifetime (Sect. \ref{introduction}) of the disc and means that there will be a vertical distribution
of grain size where the largest grains settle in the disc midplane.

The disc structure and mid-IR flux is almost fully determined by the small grains while the cool midplane, consisting of large grains,
mainly contributes to the long wavelength part of the SED and the total mass of the disc.
We construct an inhomogeneous 2-component model where the near- and mid-IR flux comes from the
small grains to which we add a blackbody flux to represent the midplane.

Stellar input parameters of the model are luminosity, mass (which we take
fixed at $L=3000\,$L$_{\odot}$ and $M=1\,$M$_{\odot}$) and T$_{eff}$.
Input disc parameters are $R_{out}$ and $R_{in}$, 
where the dust sublimation radius is used as a zero-order approximation of $R_{in}$, the total disc mass and the powerlaw
for the density distribution. For the dust sublimation temperature for silicates we use the typical value T$_{sub}=1500$\,K
and we assume blackbody radiation.
For RU\,Cen and AC\,Her this gives $R_{sub}\approx 2$\,AU. In the more detailed modelling we take the inner radius of the disc
so that the start of the IR-excess fits the photometric data.
Since we are not dealing with outflow sources a powerlaw $>-2$ is used.
Using this disc model, the SED can now be calculated, given a specific inclination angle of the system.

\subsection{Results}
\label{results}

\begin{figure}
\vspace{0cm}
\hspace{0cm}
\resizebox{9cm}{!}{ \includegraphics{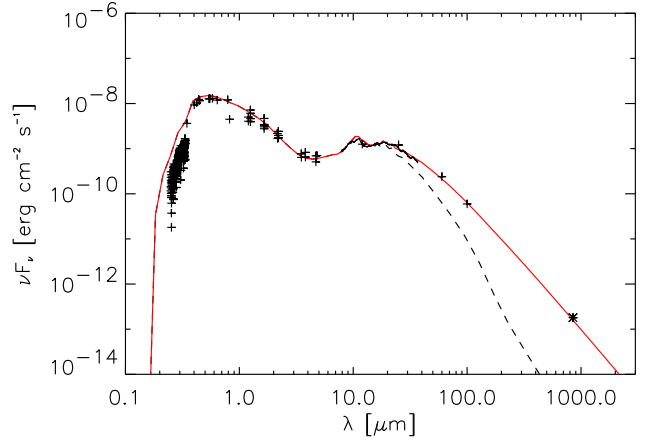}}
\caption{SED disc modelling of RU\,Cen. The dashed line represents the homogeneous disc model consisting of
grains between 0.1\,$\mu$m and 20\,$\mu$m. The red line gives the disc model with an added blackbody to represent 
the cool midplane. Crosses represent photometric data and the solid black line the observed Spitzer spectrum.
Note that the 850\,$\mu$m photometric point (asterisk) is an estimation.}
\label{SEDrucen}%
\end{figure}

\begin{figure}
\vspace{0cm}
\hspace{0cm}
\resizebox{9cm}{!}{ \includegraphics{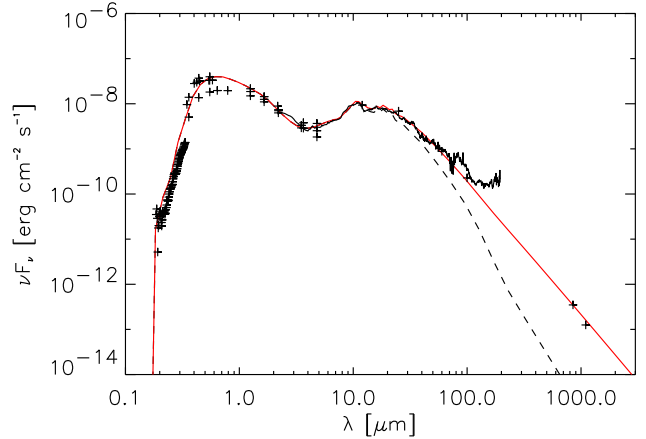}}
\caption{SED disc modelling of AC\,Her. The dashed line represents the homogeneous disc model consisting of
grains between 0.1\,$\mu$m and 20\,$\mu$m. The red line gives the disc model with an added blackbody to represent 
the cool midplane. Crosses represent photometric data and the solid black line the observed ISO-SWS spectrum.}
\label{SEDacher}%
\end{figure}

The SED-fitting gives an estimate of the distance to the systems, $d=2.4$\,kpc for
RU\,Cen and $d=1.4$\,kpc for AC\,Her, using a luminosity $L=3000\,$L$_{\odot}$.
This distance estimation is largely dependent on luminosity of
the star and adopted inclination of the system. Independent distances estimated from the P-L relation of \citet{alcock98}
are $d=1.6\pm0.6$\,kpc for RU\,Cen and $d=1.3\pm0.4$\,kpc for AC\,Her \citep{deruyter06}.
For RU\,Cen this could mean that the adopted luminosity of $L=3000\,$L$_{\odot}$
is too high.

When modelling the near- and mid-IR part of the SED the feature-to-continuum ratio
of the silicate features is too strong in comparison with the ratio observed in the infrared spectra.
Including an extra continuum opacity source is needed to reduce the strength of the features.
Since the silicates are devoid of iron (Sect. \ref{feat_ident}), metallic iron is a potential opacity source:
while its near-IR opacity is large, the absorption coefficient is unfortunately featureless so direct detection is difficult.
Inclusion of metallic iron has a strong impact on the modelling because the near-IR excess increases significantly
with a given inner radius. The opacity of metallic iron alone would require an inner radius $>70\,$AU to maintain a 
reasonable fit to the SED, but this is inconsistent with the constraints from high spatial resolution imaging \cite{close03}.
So relatively large grains (up to 20\,$\mu$m) were included in the modelling as well.

Our final model of the near- and mid-infrared part of the SED with a homogeneous disc model consists of grains
with sizes between 0.1\,$\mu$m and 20\,$\mu$m. 
In a disc, grain evolution can produce larger grains than
what is typically used for ISM grains. A grain size distribution $\sim a^p$ ($a$ is the grain size) with a powerlaw $3.5<p<-2.5$ is expected \citep{dalessio01}.
\citet{bouwman03} for example find for the Herbig\,Be star HD\,100546, a grain size distribution
$p \sim -2$ and $p \sim -2.8$ for the comet Hale-Bopp. We take $p = -3.0$ for the full range of grain sizes between 0.1
and 20\,$\mu$m to still have a significant component of small grains.
Further disc input parameters are adopted dust species, for which we take a mixture of 6\% metallic iron and 94\% amorphous and crystalline silicates.
Best models for the SED-fit of RU\,Cen and AC\,Her are presented in Fig. \ref{SEDrucen} and Fig. \ref{SEDacher}. 
For RU\,Cen we find a model with $R_{in}=35$\,AU, $R_{out}=400$\,AU and a total disc mass (in small grains and gas) of $3\times10^{-4}\,$M$_{\odot}$.
AC\,Her has a disc model with $R_{in}=35$\,AU, $R_{out}=300$\,AU and a total disc mass (in small grains and gas) of $5\times10^{-4}\,$M$_{\odot}$.
For both stars we find that the best fit is obtained when using a rather flat surface density distribution ($\Sigma \sim R^{\alpha}$, with $\alpha > -1.0$).
To reduce the number of free parameters we kept the value of this powerlaw fixed at $-0.5$.
The inclusion of both metallic iron and larger grains cause a degeneracy in the value for the inner radius. 
If one would increase the fraction of large grains, 
as main contributors to the continuum opacity, the inner radius could have values between 15\,AU and 30\,AU.
This is still significantly larger than the sublimation radius for these stars (Sect. \ref{discmodel}). 
The larger inner radius could be an evolutionary effect of the disc. Interferometric measurements are clearly needed
to further constrain the disc radii.
The added blackbody to represent the midplane and explain the far-IR part of the SED has a temperature of 120\,K for RU\,Cen and 170\,K for AC\,Her.

These temperatures and the 850\,$\mu$m fluxes can be used to estimate
the dust mass in these large grains. In the optical thin approach (at
850\,$\mu$m) the disc mass can be estimated by using \citep{hildebrand83} 
$$M_d=\frac{F_{850}\,D^2}{\kappa_{850}\,B_{850}(T)}.$$ 
Assuming a cross section of large spherical grains, the mass absorption coefficient $\kappa=\frac{\pi a^2}{\frac{4}{3}\pi a^3 \rho}$,
with $a$ the grain size and $\rho$ typically 3.3\,g\,cm$^{-3}$ for astronomical silicate, of
850\,$\mu$m grains in blackbody approximation is about 2.4\,cm$^2$\,g$^{-1}$.   
This results in dust mass estimates of $5\times 10^{-4}\,$M$_{\odot}$
for RU\,Cen and $2\times 10^{-4}\,$M$_{\odot}$ for AC\,Her.

\begin{figure}
\vspace{0cm}
\hspace{0cm}
\resizebox{8cm}{!}{ \includegraphics{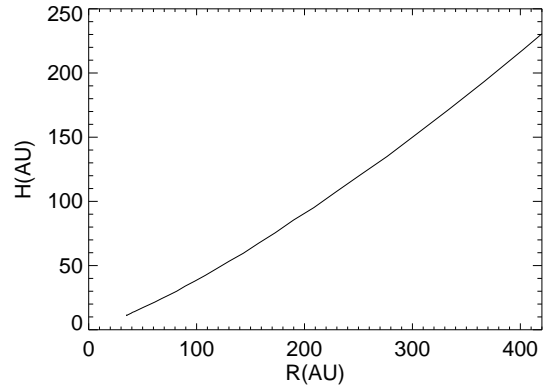}}
\caption{The scaleheight - radius relation $H\approx 0.15\times R^{1.2}$, found for the discs of RU\,Cen and AC\,Her.}
\label{hvsr}%
\end{figure}

The resulting discs for RU\,Cen and AC\,Her are flared discs, with the scale height $H\approx 0.15\times R^{1.2}$ (Fig. \ref{hvsr}).
This enables the discs to reprocess the large fraction, 15\% for RU\,Cen and 35\% for AC\,Her (Sect. \ref{sed}), of light emitted by the central star.

The models for RU\,Cen and AC\,Her have comparable discs, with similar geometries,
scale heights, temperature distributions and dust grain sizes. Temperature distributions in the discs 
are in agreement with dust temperatures derived from 
the spectral fitting. Both discs are optically thick in the equator
direction between 0.1\,$\mu$m and 1\,$\mu$m. After 1\,$\mu$m the optical thickness decreases rapidly,
making the disc optically thin for infrared emission.

\section{Discussion}
\label{discussion}

A major problem in the SED model fitting of the the near- and mid-infrared flux is that the models are  
degenerate, especially the outer radius and the total disc mass are poorly constrained. 
Almost equally fitting models for RU\,Cen and AC\,Her can be found
with an outer radius a few hundred AU larger which would result in
a larger total disc mass. It is also difficult to discriminate between
the two opacity sources we added to reduce the
feature-to-continuum ratio of the silicate features: metallic iron and
larger grains. If grains up to 50\,$\mu$m are included in the disc, the amount of metallic
iron needed to fit the spectrum will be reduced significantly.

To compare the modelled SED with observational photometric data, these data 
need to be corrected for the interstellar extinction. A value for the interstellar extinction would constrain the inclination
of the system. Since no such estimate is available we put a minimum on the interstellar extinction of $E(B-V)_{is}=0.1$. 
As a maximum value we use the total extinction which we have deduced in Sect. \ref{sed}.
This gives us possible inclination values smaller than $70^{\circ}$ for RU\,Cen and smaller than $50^{\circ}$ for AC\,Her.

With these values for the inclination a minimal mass for the companion star can be estimated, using the mass function
and a typical value for the primary of $M_1=0.5-0.6$\,M$_{\odot}$.
The mass functions of $f(M)=0.83$\,M$_{\odot}$ for RU\,Cen and  $f(M)=0.25$\,M$_{\odot}$ for AC\,Her \citep{vanwinckel98} 
yield a minimal mass for the companion star of
$M_2=1.7\,$M$_{\odot}$ for RU\,Cen and $M_2=1.1\,$M$_{\odot}$
for AC\,Her. Even the minimal mass of the companion of RU\,Cen is therefore
not compliant with a possible white dwarf mass, and the companions are
in both cases likely to be unevolved main sequence stars.

\section{Conclusion}
\label{conclusion}

AC\,Her and RU\,Cen are known to be proto-typical RV\,Tauri pulsators
which are normally seen as transition objects in their evolution from
the AGB to the PNe evolutionary phase. Furthermore RU\,Cen turned out to be an evolved binary with an orbital
period similar to AC\,Her indicating that it must have been subject to severe binary
interaction when at giant dimensions. Neither of the two currently fills its
Roche Lobe. In this paper we focused on modelling the
circumstellar environment as constrained by our high quality Spitzer
spectra and the broad-band SED. Our analysis showed that both stars
are surrounded by a circumbinary dusty 
disc in hydrostatic equilibrium. Since the disc modelling is a
degenerate problem, interferometric measurements are needed to
constrain further the disc geometry (e.g. \citealt{deroo07b,deroo07a}).

The observed Spitzer spectra clearly show that the circumstellar
grains are extremely processed. The IR-spectrum of both objects is
dominated by crystalline dust features. The mineralogy is
magnesium rich and large grains and/or metallic iron is necessary to
explain the low feature-to-continuum ratio of the silicate
features. The temperature estimate of the crystalline silicates show
that a significant fraction must be rather cool, and certainly well below the annealing
temperature. This shows that either radial mixing from the hot inner
boundary (where annealing can take place) to far out in the disc must have
occurred or that the formation process and thermal history of the grains
is quite different in discs than in outflows. The profile fitting
shows that the grains must be very irregular.

Despite the very different evolutionary history and the very different
evolutionary timescales involved, it is remarkable that the mineralogy of
the small hot silicates around the evolved objects RU\,Cen and AC\,Her is extremely
similar to what is found in some young stellar objects (YSO) such as HD100546 \citep{malfait98,lisse07} and that
of primitive comets such as Hale-Bopp \citep{bouwman03,lisse07,min05b}.
In all those systems, the thermal history of the grains has been such as to
promote the dominance of forsterite, the Mg-rich end member of the crystalline
olivine family. In YSO as well as in comets, the dust
processing is thought to be the tracer for the process of disc clearing and
planet building in the proto-planetary disc.

In evolved objects the circumstellar material is coming from the stars
themselves and is much more chemically homogeneous than the ISM composition
around YSO. Moreover the evolutionary timescales are
likely to be orders of magnitude smaller than the disc evolution in YSO. Detailed dust
mineralogy studies around evolved stars can therefore yield important
information on dust formation and dust processing in discs, and this under very
different chemical, dynamical and evolutionary environments than the processing
in proto-planetary discs. Our study appears to indicate that the
chemico-physical processes of dust grains in the hydrostatic discs of the
evolved binaries RU\,Cen and AC\,Her, are very similar to those governing in
the protoplanetary discs around YSO. We are in the process of expanding our
study to a wider sample of evolved binaries. This will enable us to build up a
broader view of the chemico-physical dust processes of the grains around rapidly
evolving stars.

The inclination and mass function give a minimal
mass for the companion star of $1.1\,$M$_{\odot}$ for AC\,Her and
$1.7\,$M$_{\odot}$ for RU\,Cen.
The unevolved companions have masses such that they would normally (on single-star evolutionary
tracks) evolve to carbon stars on the AGB. Given the silicate
dominated circumstellar dust, it is however, clear that even the primaries did not
evolve to become carbon stars. The actual orbits combined with the
observed chemical evolution of the stars show that the binary
interaction processes dominate the final evolution of these
objects. The internal chemical evolution seems to have been cut short by binary
interaction processes. As such the stars do not evolve on single-star evolutionary
tracks and both objects should be seen as remarkably evolved
binaries. It is likely that the formation of the circumstellar dust in
these objects is closely related to binary interaction.

Both objects show that the formation, structure and evolution of the
circumstellar discs play a leading role in their final evolution and
illustrate once again the intimate relation between the depleted
photospheres and the presence of a circumbinary disc. Since
post-AGB stars with similar SEDs and/or chemistry are abundant
(e.g. \citealt{deruyter06,reyniers07}) it is clear that disc formation is a
process relevant in the late evolution of a considerable fraction of binary stars.

\begin{acknowledgements}
The authors want to acknowledge: the Geneva Observatory and its staff for the
generous time allocation on the Swiss Euler telescope; the 1.2\,m Mercator staff
as well as the observers from the Instituut voor Sterrenkunde who contributed
to the monitoring observations using both the Euler and Mercator telescopes.
CG acknowledges support of the Fund for Scientific Research of Flanders
(FWO) under the grant G.0178.02. and G.0470.07.
We also thank Fred Lahuis for his assistance with the Spitzer data reduction.
\end{acknowledgements}

\bibliographystyle{aa}
\bibliography{8323bib}
\end{document}